\documentclass[5p]{elsarticle}%
\usepackage{amsfonts}
\usepackage{amsmath}
\usepackage{graphicx}
\usepackage{amssymb}
\usepackage{caption}%
\providecommand{\U}[1]{\protect\rule{.1in}{.1in}}
\begin{document}
\begin{frontmatter}
\title{On the frequency of oscillations in the pair plasma generated by a strong electric field}
\author{A. Benedetti}
\ead{Alberto.Benedetti@icra.it}
\author{W.-B. Han}
\ead{wenbiao@icra.it}
\author{R. Ruffini}
\ead{ruffini@icra.it}
\author{G.V. Vereshchagin}
\ead{veresh@icra.it}
\address{ICRANet, P.le della Repubblica 10, 65100 Pescara, Italy, \\
ICRA and University of Rome "Sapienza", Physics Department, \\
P.le A. Moro 5, 00185 Rome, Italy.}
\begin{abstract}
We study the frequency of the plasma oscillations of electron-positron pairs created by the vacuum polarization in an uniform electric field with strength $E$ in the range $0.2\,E_{c}<E<10\,E_{c}$. Following the approach adopted in \cite{VacuumPol} we work out one second order ordinary differential equation for a variable related to the velocity from which we can recover the classical plasma oscillation equation when $E\rightarrow 0$. Thereby, we focus our attention on its evolution in time studying how this oscillation frequency approaches the plasma frequency. The time-scale needed to approach to the plasma frequency and the power spectrum of these oscillations are computed. The characteristic frequency of the power spectrum is determined uniquely from the initial value of the electric field strength. The effects of plasma degeneracy and pair annihilation are discussed.
\end{abstract}
\end{frontmatter}





The electron-positron pair production in a strong electric field is one of the
most popular topics in relativistic field theory \cite{PairPhysAstro}. It
begun with the pioneer works by Sauter \cite{S}, Heisenberg and Euler
\cite{S1}, and by Schwinger \cite{S2}. This effect acquires particular
importance when the electric field strength $E$ is larger than the critical
value $E_{c}\equiv m^{2}c^{3}/(e\hbar)$; such a strong electric field can be
reached in astrophysical environments, near quark stars \cite{Usov}%
-\cite{Usov2} and neutron stars \cite{NS1}-\cite{NS2}. Strong electric fields
up to several percents of the critical value will be reached by
advanced laser technologies in laboratory experiments \cite{Ringwald}%
-\cite{Gorodienko}, X-ray free electron laser facilities \cite{XFEL}, optical
high-intensity laser facilities such as Vulcan or the Extreme Light
Infrastructure \cite{ELI}, for a recent review see \cite{UFN}. Electron
beam-laser interactions seem also promising in reaching high Lorentz
transformed electromagnetic fields capable for multiple pair production
\cite{Sokolov}.

It has been shown that, due to back reaction and screening effects of
$e^{+}e^{-}$ pairs on external electric fields, positrons and electrons move
back and forth coherently with alternating electric field: the so called
plasma oscillations. In \cite{VacuumPol} it was pointed out that this
phenomenon occurs also when $E\leq E_{c}$ giving emphasis on the fact that,
for overcritical (undercritical) field, a large (small) fraction of the
initial electromagnetic energy is converted into the rest mass of pairs,
whereas a small (large) fraction is converted into kinetic energy. In
\cite{wenbiao} the case of spatially inhomogeneous electric field has been
considered, the emitted radiation spectrum far from the oscillation region was
obtained, presenting a narrow feature.

In this Letter we return to basic equations describing pair creation and
plasma oscillations in uniform unbound electric field. We first derive a
master equation for a new variable constructed from hydrodynamic velocity,
which turns out to be second order ordinary differential equation. This
equation is reduced to the classic plasma oscillations equation describing
Langmuir waves in the limit of small electric field. The frequency of
oscillations is then shown to be almost equal to the plasma frequency, which
is strongly time dependent in the case under consideration. Finally, the
spectrum of bremsstrahlung radiation is computed following \cite{wenbiao} and
its characteristic feature is identified as a function of initial value of the
electric field strength.

As in \cite{VacuumPol} we apply an approach based on continuity,
energy-momentum conservation and Maxwell equations in order to account for the
back reaction of the created pairs focusing on the range $0.2\,E_{c}%
<E<10\,E_{c}$.

We assume that electrons and positrons are created at rest in pairs, due to
vacuum polarization in uniform electric field \cite{S}-\cite{S2},
\cite{S3}-\cite{book2} with the average rate\footnote{We use in the following
the system of units where $\hbar=c=1$, $e=\sqrt{\alpha}\approx\sqrt{1/137}$,
$\alpha$ being the fine structure constant.} per unit volume $V$ and per unit
time $t$
\begin{align}
S\equiv\frac{dN}{dVdt}=\frac{m^{4}}{4\pi^{3}}\left(  \frac{E}{E_{c}}\right)
^{2}\exp\left(  -\pi\frac{E_{c}}{E}\right)  ,\label{rate}\\
E=\sqrt{-\frac{1}{2}F_{\mu\nu}F^{\mu\nu}},
\end{align}
where $F^{\mu\nu}$ is electromagnetic field tensor, $m$\ is electron mass.

This formula is derived for uniform constant in time electric field. However,
it still can be used for slowly time-varying electric field provided the
inverse adiabaticity parameter \cite{book1}-\cite{Popov} is much larger than
one,
\begin{equation}
\eta=\frac{m}{\omega}\frac{E_{peak}}{E_{c}}=\tilde{T}\tilde{E}_{peak}\gg1,
\label{eta}%
\end{equation}
where $\omega$ is the frequency of oscillations, $\tilde{T}=m/\omega$ is
dimensionless period of oscillations. Eq. (\ref{eta}) implies that time
variation of the electric field is much slower than the rate of pair
production. In two limiting cases considered in this Letter, $E_{10}=10E_{c}$
and $E_{0.2}=0.2E_{c}$, we find respectively for the first oscillation
$\eta_{10}=334$, and\thinspace$\eta_{0.2}=4.8\times10^{5}$, whereas for the
last one $\eta_{10}=1.5$, and $\eta_{0.2}=4$.

Following \cite{VacuumPol} the conservation laws and Maxwell equations written
for electrons, positrons and electromagnetic field are
\begin{align}
\frac{\partial\left(  \bar{n}U^{\mu}\right)  }{\partial x^{\mu}}  &
=S,\label{cont}\\
\frac{\partial T^{\mu\nu}}{\partial x^{\nu}}  &  =-F^{\mu\nu}J_{\nu
},\label{em}\\
\frac{\partial F^{\mu\nu}}{\partial x^{\nu}}  &  =-4\pi J^{\mu}, \label{me}%
\end{align}
where $\bar{n}$ is the comoving number density of electrons, $T^{\mu\nu}$ is
energy-momentum tensor of electrons and positrons
\begin{equation}
T^{\mu\nu}=m\bar{n}\left(  U_{(+)}^{\mu}U_{(+)}^{\nu}+U_{(-)}^{\mu}%
U_{(-)}^{\nu}\right)  , \label{emten}%
\end{equation}
where $J^{\mu}=J_{cond}^{\mu}+J_{pol}^{\mu}$ is the total four-current
density, $U^{\mu}$ is four velocity respectively of positrons and electrons.
Electrons and positrons move along the electric field lines in opposite directions.

It has been shown in \cite{VacuumPol} that in a uniform electric field, from
the system (\ref{cont})-(\ref{me}), the following system of four coupled
ordinary differential equations may be obtained
\begin{align}
\frac{d\tilde{n}}{d\tilde{t}} &  =\tilde{S},\label{ratesystem}\\
\frac{d\tilde{\rho}}{d\tilde{t}} &  =\tilde{n}\tilde{E}\tilde{v}+\tilde
{\gamma}\tilde{S},\label{numsys}\\
\frac{d\tilde{p}}{d\tilde{t}} &  =\tilde{n}\tilde{E}+\tilde{\gamma}\tilde
{v}\tilde{S},\label{momentumsystem}\\
\frac{d\tilde{E}}{d\tilde{t}} &  =-8\pi\alpha\left(  \tilde{n}\tilde{v}%
+\frac{\tilde{\gamma}\tilde{S}}{\tilde{E}}\right)  ,\label{Edot}%
\end{align}
where $n=m^{3}\tilde{n}$ is dimensionless number density normalized by the Compton length $\lambda_c=1/m$, $\rho=m^{4}%
\tilde{\rho}$ is energy density of positrons\footnote{Total energy density of
electrons and positrons is twice this value.}, $p=m^{4}\tilde{p}$\ is momentum
density of positrons, $E=E_{c}\tilde{E}$\ is electric field strength, and
$t=m^{-1}\tilde{t}$\ is\ time, normalized by the Compton time $t_c=1/m$. The rate of pair production is $\tilde{S}%
=\frac{1}{4\pi^{3}}\tilde{E}^{2}\exp\left(  -\frac{\pi}{\tilde{E}}\right)  $,
velocity is $\tilde{v}=\tilde{p}/\tilde{\rho}$ and Lorentz factor is
$\tilde{\gamma}=\left(  1-\tilde{v}^{2}\right)  ^{-1/2}$. \begin{figure}[pth]
\begin{center}
\includegraphics[width=3in]{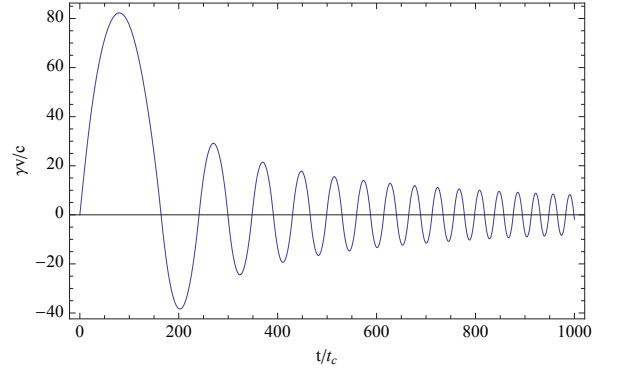}
\end{center}
\caption{Numerical solution of (\ref{ddueq}) for $E_{0}=2E_{c}$. This figure
shows damped oscillations with a frequency increasing in time.}%
\label{u_log_global_2}%
\end{figure}\begin{figure}[pth]
\begin{center}
\includegraphics[width=3in]{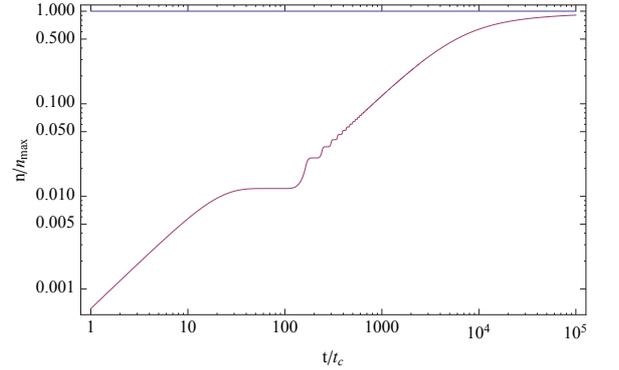}
\end{center}
\caption{Ratio between number density and maximum achievable number density
$n_{max}=1/(4\pi\alpha)\;\lambda_{c}^{-3}$ for $E_{0}=2E_{c}$; it becomes
close to unity after $10^{5}\,t_{c}$. From Fig. \ref{u_log_global_2} and Eq.
(\ref{eqm}) we see that the maxima of $\tilde{u}$\ correspond to $E=0$ and
quenching of pair creation giving rise to flattening of $n$. This happens for
each oscillation, but it is more evident at the beginning due to the double
logarithmic scale on this figure.}%
\label{density_2}%
\end{figure}

\begin{figure}[pbh]
\begin{center}
\includegraphics[width=3in]{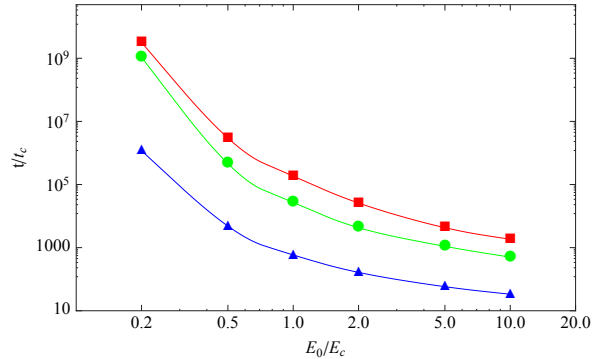}
\end{center}
\caption{In this figure are shown, for all the cases under interest, the half
period of the first oscillation $t_{1}$ (triangles), the characteristic time-scales $t_a$ (squares) needed to the pairs oscillation frequency to reach the plasma frequency and the time $t_\gamma$ (circles) which satisfies the condition $\tau(t_\gamma)\simeq1$ for the optical depth defined in (\ref{tau}) . The value of $t_a$ for the case $E_{0}=2E_{c}$ is shown in fig. \ref{ratio_2} by the
vertical line.}%
\label{3times_joined}%
\end{figure}

For the system (\ref{ratesystem})-(\ref{Edot}) there exist two integrals
(conservation laws)%
\begin{align}
{\tilde{\rho}}^{2} &  ={\tilde{p}}^{2}+{\tilde{n}}^{2}\label{energy1}\\
16\pi\alpha\tilde{\rho} &  ={\tilde{E}_{0}}^{2}-{\tilde{E}}^{2},\label{energy}%
\end{align}
so the particle energy density vanishes for initial value of the electric
field $\tilde{E}_{0}$. Combining together the previous two equations, we get
for the maximum number density of pairs that can be created
\begin{equation}
\tilde{n}_{max}=\frac{\tilde{E_{0}}^{2}}{16\pi\alpha}.\label{numdensmax}%
\end{equation}
In \cite{VacuumPol} the system (\ref{ratesystem})-(\ref{Edot}) was reduced to
two equations using (\ref{energy1})-(\ref{energy}) and analyzed on the phase
plane ($\tilde{E},\tilde{v}$).

Notice that the equation of motion of single particle in our approximation is
just%
\begin{equation}
\dot{\tilde{u}}=\tilde{E}, \label{eqm}%
\end{equation}
where we have defined $\dot{\tilde{u}}=d\tilde{u}/d\tilde{t}$ and introduced a
new variable constructed from hydrodynamic velocity as $\tilde{u}%
=\tilde{\gamma}\tilde{v}=\gamma v/c$. Then this equation can be combined with
(\ref{Edot}) to obtain a single master equation
\begin{equation}
\ddot{\tilde{u}}+\frac{{\tilde{E}_{0}}^{2}-{\dot{\tilde{u}}^{2}}}%
{2(1+\tilde{u}^{2})}\tilde{u}+\frac{2\alpha}{\pi^{2}}\left(  1+{\tilde{u}}%
^{2}\right)  ^{1/2}\exp{\left(  -\frac{\pi}{|\dot{\tilde{u}}|}\right)
\dot{\tilde{u}}=0}. \label{ddueq}%
\end{equation}
The key point of our treatment is the physical interpretation of Eq. (\ref{ddueq}) which can be rewritten symbolically as%
\begin{equation}
\ddot{\tilde{u}}+\tilde{\omega}_{p}^{2}\,\tilde{u}+k\,\dot{\tilde{u}}=0.
\label{harmoscill}%
\end{equation}
With constant coefficients Eq. (\ref{harmoscill}) would describe damped
harmonic oscillations with frequency ${\tilde{\omega}}_{p}$ and friction $k$.
In our case $\tilde{\omega}_{p}$ and $k$ are time dependent, but Eq.
(\ref{ddueq}) still possesses an oscillating behavior with damping. With our
definitions the number density of pairs is
\begin{equation}
\tilde{n}=\frac{\tilde{E_{0}}^{2}-{\dot{\tilde{u}}}^{2}}{16\pi\alpha
(1+\tilde{u}^{2})^{1/2}}. \label{numdens}%
\end{equation}
We then identify ${\tilde{\omega}}_{p}$ in (\ref{harmoscill}) as
\begin{equation}
{\tilde{\omega}}_{p}=\sqrt{\frac{8\pi\alpha\,\tilde{n}}{(1+\tilde{u}%
^{2})^{1/2}}}, \label{plasmafreq}%
\end{equation}
i.e. the relativistic plasma frequency\footnote{The factor $8\pi$ is in this
formula due to the presence of two charge carriers with the same mass -
electrons and positrons. This is different from the classical electron-ion
plasma where only electron component oscillates and the corresponding factor
is twice smaller.}.

The function
\begin{equation}
k=\frac{2\alpha}{\pi^{2}}\left(  1+{\tilde{u}}^{2}\right)  ^{1/2}\exp{\left(
-\frac{\pi}{|\dot{\tilde{u}}|}\right)  } \label{kappa}%
\end{equation}
in Eq. (\ref{harmoscill}) accounts for the rate of pair production
(\ref{rate}). It describes the increase of inertia of electron-positron pairs
due to increase of their number and causes decrease of the amplitude of oscillations.

For small electric fields Eq. (\ref{ddueq}) is reduced to classical plasma
oscillations equation describing Langmuir waves, since in that case $k$ is
exponentially suppressed. For this reason we expect that as the amplitude of
oscillations of electric field gets smaller the frequency of oscillations
$\omega$ tends to the plasma frequency $\omega_{p}$.

In \cite{VacuumPol} we solved numerically the system of four coupled ordinary
differential equations (\ref{ratesystem})-(\ref{Edot}). Now it is possible to
solve just one second order differential equation (\ref{ddueq}) which allows
us to study its asymptotic behavior as well.

We solve numerically Eq. (\ref{ddueq}) with the initial conditions
$\dot{\tilde{u}}(0)=\tilde{E}_{0}$ and $\tilde{u}(0)=0$, corresponding to no
pairs in the initial moment, taking for initial electric field strength
$\tilde{E}_{0}=\{0.2,0.5,1,2,5,10\}$. Once this equation has been solved, we
have the solution for the number density from (\ref{numdens}) and for the
plasma frequency by means of (\ref{plasmafreq}). In Fig. \ref{u_log_global_2}
we show the evolution in time of $\tilde{u}$ where the amplitude of the
oscillations decreases, while its frequency increases with time.

We present the number density of pairs in Fig. \ref{density_2} for the case
$E_{0}=2\,E_{c}$ as a fraction of the maximum achievable value $\tilde
{n}_{\max}$. In all the cases under interest, this number is asymptotically
achieved indicating that the final result of the process will be the complete
conversion of the electromagnetic energy density into the rest mass of the
pairs. Moreover, looking at Fig. \ref{density_2} we recover the result
obtained in \cite{VacuumPol}; in fact we can see that after the first
oscillation, higher is $E_{0}$ larger is $\tilde{n}$. This means that the
first oscillation gives the leading contribution to the process in which the
electromagnetic energy of the field is converted in the rest mass of pairs,
with a moderate contribution to their kinetic energy for $E_{0}>E_{c}$. The
values of the half periods of the first oscillation for each considered case
are represented in Fig. \ref{3times_joined} by triangles.

We computed the frequency of the $i$-th oscillation as $\tilde{\omega}^{i}%
=\pi/\tilde{T}_{1/2}^{i}$, where $\tilde{T}_{1/2}^{i}$ is the corresponding
half period, calculated considering the time interval between the $i$-th and
($i+1$)-th subsequent roots of $\tilde{u}$, see Fig. \ref{u_log_global_2}.

\begin{figure}[pth]
\begin{center}
\includegraphics[width=3in]{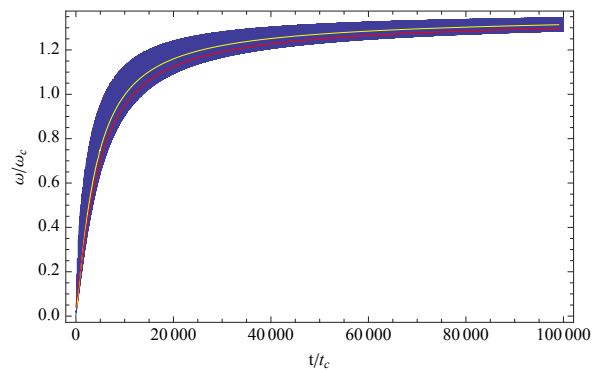}
\end{center}
\caption[width=0.8\textwidth]{In this plot the blue area represents the plasma
frequency as defined in Eq. (\ref{plasmafreq}); it appears like a continuum
because of the fast oscillations. The yellow curve is its average in time
${\tilde{\omega}}_{p}^{av}$ which can be compared with the pairs oscillation
frequency $\tilde{\omega}$ given by the red curve. This plot corresponds to
the case $E_{0}=2E_{c}$.}%
\label{freq_comparison_2}%
\end{figure}

\begin{figure}[pth]
\begin{center}
\includegraphics[width=3in]{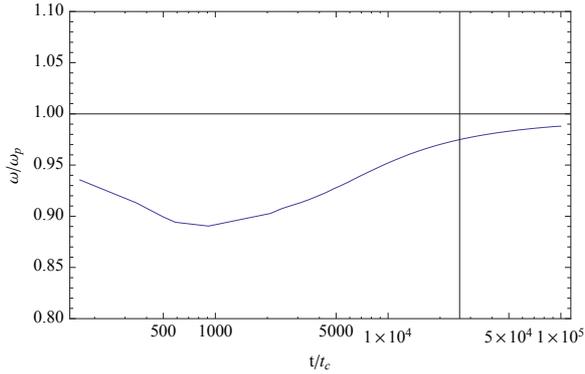}
\end{center}
\caption{Behavior of the ratio ${\tilde{\omega}}/{\tilde{\omega}}_{p}^{av}$ in
time for $E_{0}=2E_{c}$. The plasma frequency is attained asymptotically
because of the limit $\tilde{S}(\tilde{t}\rightarrow\infty)\rightarrow0$. The
vertical line corresponds to the time-scale needed to attain the plasma
frequency in this specific case.}%
\label{ratio_2}%
\end{figure}

Notice that ${\tilde{\omega}}_{p}$ is an oscillating function of time, due to
the presence of ${\tilde{u}}$ in (\ref{plasmafreq}); besides the frequency of
these oscillations increases in time. For this reason, in order to get a
smooth function we calculated the average of ${\tilde{\omega}}_{p}$. We use
this new function ${\tilde{\omega}}_{p}^{av}$ for the plasma frequency to make
a comparison with the frequency of oscillations of pairs. In Fig.
\ref{freq_comparison_2}, for the case $E_{0}=2E_{c}$, the blue area represents
the plasma frequency as defined by (\ref{plasmafreq}), the yellow curve is its
average ${\tilde{\omega}}_{p}^{av}$, while the pairs oscillation frequency
$\tilde{\omega}$ is represented by the red curve. For the same initial
electric field, in Fig. \ref{ratio_2} the trend of the ratio $\tilde{\omega
}/{\tilde{\omega}}_{p}^{av}$ is shown, which indicates that the averaged
plasma frequency is achieved asymptotically as expected from (\ref{ddueq}).
Notice that the oscillation frequency $\tilde{\omega}$ is always smaller than
${\tilde{\omega}}_{p}^{av}$ since the number density of pairs is constantly
increasing with time during each oscillation cycle.

\begin{figure}[pth]
\begin{center}
\includegraphics[width=3in]{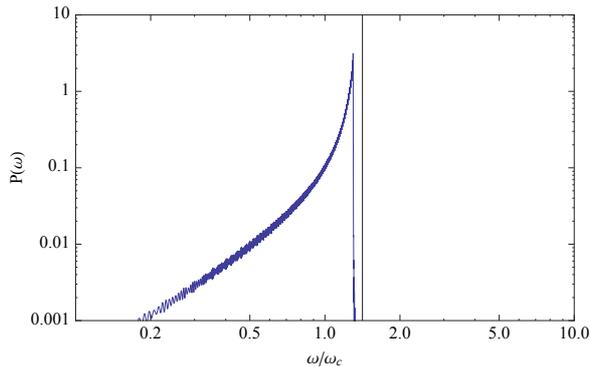}
\end{center}
\caption{The power spectrum in arbitrary units has been obtained using Eq.
(\ref{powspectrum}). The peak almost corresponds to the maximum plasma
frequency (\ref{plasmafreq}) corresponding to the maximum achievable number
density (\ref{numdensmax}), indicated here by the vertical line. The main
contribution is given by the oscillations which last for a long time, namely
when the asymptotic limit of the plasma frequency is attained.}%
\label{spectrum_2}%
\end{figure}

We computed the power spectrum of radiation in the far zone, assuming dipole
radiation following \cite{wenbiao}. The power spectrum, namely the energy
radiated per unit solid angle per frequency interval and per unit volume, is
given by
\begin{equation}
\mathbf{\tilde{P}}(\tilde{\omega})=\frac{d\tilde{I}}{d\tilde{\omega}\,d\Omega
}=2\alpha\,|\mathbf{\tilde{D}}({\tilde{\omega}})|^{2},\label{powspectrum}%
\end{equation}
where the amplitude $\tilde{\mathbf{D}}(\tilde{\omega})$ is proportional to
the Fourier transform of the electric current time derivative \cite{wenbiao}
\begin{equation}
\mathbf{\tilde{D}}({\tilde{\omega}})\propto\int_{\tilde{T}}d\tilde
{t}\,e^{i{\tilde{\omega}}\tilde{t}}\left[  \frac{\partial\mathbf{\tilde{J}%
}(\tilde{t})}{\partial\tilde{t}}\right]  .\label{Domega}%
\end{equation}
The electric current is simply related to the new variables by the following
expression
\begin{equation}
\tilde{J}=2\,\sqrt{\alpha}\tilde{n}\,\frac{\tilde{u}}{\left(  1+{\tilde{u}%
}^{2}\right)  ^{1/2}}.
\end{equation}
From the Fig. \ref{spectrum_2} it is clear that the main contribution is given
by the final and fastest oscillations which last for a longer time. Therefore,
the power spectrum shows a peak close to the plasma frequency being reached
asymptotically. We can easily estimate the frequency corresponding to this peak combining Eq. (\ref{numdensmax}) with Eq. (\ref{plasmafreq}) as
\begin{equation}
\tilde{\omega}_{peak}\simeq\frac{\tilde{E}_0}{\sqrt{2}}\label{omegapeak}
\end{equation}
with the corresponding energy $\hbar\omega_{peak}\simeq0.72\,E_0/E_c$ MeV.

The energy loss due to the dipole radiation for $E=2E_{c}$
considered in Fig. \ref{spectrum_2} for $t=10^{5}t_{c}$ is less than one
percent. 

Once we know the frequency of the pairs oscillations and the plasma frequency,
we obtain their ratio as it is shown in Fig. \ref{ratio_2}. Besides, we use
${\tilde{\omega}}/{\tilde{\omega}}_{p}^{av}$ in order to compute the
characteristic time scale $t_a$ needed for the pairs oscillation
frequency to reach the plasma frequency. This has been done considering the
ratio between ${\tilde{\omega}}/{\tilde{\omega}}_{p}^{av}$ and its time
derivative; the result of this procedure gives us a numerical function from
which we have taken the average. For all the considered cases, this quantity
is shown in Fig. \ref{3times_joined}, from which we understand that the general
trend is that larger is the initial electric field, larger will be the
starting oscillation frequency. 

It is worth noting the effect of degeneracy on the pair production. One may
think that when concentration of pairs reaches the maximum allowed value by
the Pauli principle the pair production is blocked. For particles \emph{at
rest} this would happen when two pairs with opposite spins occupy a Compton
volume. Considering asymptotic number of pairs given by (\ref{numdensmax}) one
finds that it would happen for $E>4\sqrt{\alpha\pi}E_{c}\simeq0.6E_{c}$.
However, one has to keep in mind that particles produced at rest are
accelerated by external electric field and thus leave the quantum state with
zero momentum which can be subsequently filled by a new pair. These effects
are independent since they operate in ortogonal directions of the phase space,
so one can estimate the value of external electric field at which phase space
blocking occurs by comparing their rates. Such analysis gives us the following
inequality%
\begin{equation}
\frac{1}{4\pi^{3}}\tilde{E}\exp\left(  -\frac{\pi}{\tilde{E}}\right)
\geq1,\label{blockingrate}%
\end{equation}
having the solution $E\gtrsim127E_{c}$, which is much higher than electric
fields considered in this Letter.

Another effect, relevant for large enough electric field, is interaction of
pairs with photons discussed in some details in \cite{VacuumPol,Astro}. One
can estimate the optical depth for electron-positron annihilation as%
\begin{equation}
\tau(t)\simeq\int_{0}^{t}\frac{\sigma_{T}}{\gamma^{2}}nvdt=\int_{0}^{\tilde
{t}}\frac{8\pi\alpha^{2}}{3}\frac{|\tilde{u}|}{\left(  1+{\tilde{u}}^{2}\right)
^{3/2}}\,\tilde{n}\,d\tilde{t},\label{tau}%
\end{equation}
where $\sigma_{T}$ is the Thomson's cross section, and we approximated
$\sigma\simeq\sigma_{T}/\gamma^{2}$. Equating (\ref{tau}) to unity we find the
timescale $t_\gamma$ at which the probability of electron to interact with positron and
create a pair of photons reaches unity. From that time moment interaction of
pairs with photons can no longer be neglected. This timescale is represented
in Fig. \ref{3times_joined} by circles.

Summarizing the informations presented in Fig. \ref{3times_joined} we conclude that independent on the initial value of the electric field, there is a hierarchy between the following time scales $t_1<t_\gamma<t_a$. It means that many oscillations occur before electron-positron collisions turn out to be important, thus justifying our collisionless approximation. The condition $t_\gamma<t_a$ means that the estimation of the maximal frequency of oscillations (\ref{omegapeak}) is an approximate one: photons produced by interaction of pairs will also distort the spectrum shown in Fig. \ref{spectrum_2}.

As long as electric field does not reach critical values for creation of muons
and pions their production from electron-positron collisions \cite{Muller}%
-\cite{Kuznetsova} is suppressed because of two different mechanisms. Both
these processes have a kinematic threshold given by the rest mass of the
produced particles. For this reason the Lorentz factor of the relative motion
of colliding electron and positron should exceed $\sim10^{2}$, restricting
initial electric fields to be undercritical, $E_{0}<E_{c}$, see Fig. 3 in Ref.
\cite{VacuumPol}. On the other hand, the number density of pairs produced is
exponentially suppressed for undercritical fields. Besides the cross section
for all these processes decreases as $\sigma\propto\gamma^{-2}$ which further
decreases the rate of electron-positron collisions.

It has been claimed recently \cite{Fetodov} that critical Schwinger field
could never be reached in high power lasers due to occurence of avalanche-like
QED cascade operating mainly through nonlinear Compton scattering combined
with nonlinear Breit-Wheeler process \cite{BellKirk,KirkBell}, see also
\cite{PairPhysAstro}, and via the trident process \cite{hu,Ilderton}. As soon as
one single pair is generated by the Schwinger process such electromagnetic
cascade of secondary electron-positron pairs is expected to deplete the
electromagnetic energy thus preventing further pair production from vacuum.
The requirements for the avalanche to occur are twofold: a)\ the probability
to emit photon should not be suppressed and b) the photon must be energetic
enough to produce pair by interaction with another photon. It is shown that
for a specific electromagnetic field configuration considered in
\cite{BellKirk,KirkBell} as well as in \cite{Fetodov}, namely circularly
polarized standing electromagnetic wave both these conditions may fulfill for
undercritical electric field $E<E_{c}$. However, as it was shown in
\cite{Bulanov} for linearly polarized standing wave such electromagnetic
cascade is not expected to dominate over the Schwinger process. It is easy to
understand these results looking at the energy loss rate of charged particle
in classical electrodynamics
\begin{equation}
\frac{dW}{dt}=\frac{2}{3}\frac{\alpha^{2}}{m^{2}}\gamma^{2}\left[  \left(
\mathbf{E}+\mathbf{v}\times\mathbf{H}\right)  ^{2}-\left(  \mathbf{E}%
\cdot\mathbf{v}\right)  ^{2}\right]  .\label{energyloss}%
\end{equation}
When magnetic field is absent (the case considered in \cite{BellKirk,KirkBell}
and \cite{Fetodov}) if directions of particle velocities and electric field
are collinear the radiation loss turns out independent on particle energy. In
such case, as we have shown previously \cite{VacuumPol} for overcritical
electric field the radiation loss is smaller than the rate of energy
conversion from electromagnetic field to electron-positron pairs via Schwinger
process. Notice that in the case of plasma oscillations considered in this
Letter the velocity and acceleration vectors are indeed collinear, so
curvature radiation considered in \cite{BellKirk,KirkBell} does not occur.
When electric field changes with time not only its amplitude but also
direction, as for instance in circularly polarized electromagnetic wave, the
acceleration and velocity vectors become misaligned and curvature radiation
becomes much more efficient due to quadratic dependence on particle energy in
(\ref{energyloss}). We also notice that the backreaction of electron-positron
pairs on the initial electric field, which is the topic of the present Letter,
is not taken into account in \cite{BellKirk,KirkBell} and \cite{Fetodov}, see however \cite{nerush}.

To summarize, the study of plasma oscillations due to the vacuum polarization
in uniform electric field can be reduced to the analysis of a single second
order ordinary differential equation for the variable constructed from
hydrodynamic velocity $\tilde{u}=\gamma v/c$. All the other physical
quantities of interest can be obtained from the solution of (\ref{ddueq}).
This reduction allows to study the evolution of the system for a long time.

As expected, the plasma frequency is reached asymptotically for all the
considered cases $0.2\,E_{c}\leq E_{0}\leq10\,E_{c}$. The difference between
them rely on the time scale the system needs to approach the plasma frequency
as it is shown in Fig. \ref{3times_joined}. In particular, larger is the initial
electric field, shorter will be the time scale to get the plasma frequency.

Surprisingly we find that, for all the cases we have considered,
${\tilde{\omega}}\simeq{\tilde{\omega}}_{p}^{av}$ even for the very first
oscillations, when we are far from the asymptotic case $\tilde{S}(\tilde
{t}\rightarrow\infty)\rightarrow0$ when we expect ${\tilde{\omega}}%
={\tilde{\omega}}_{p}^{av}$ from the analysis of Eq. (\ref{ddueq}).

The characteristic feature of the power spectrum of dipole radiation occuring
due to plasma oscillations is shown to be located close, but always below, to
the plasma frequency. The left tail in Fig. \ref{spectrum_2} is due to the
first oscillations with frequencies smaller than ${\tilde{\omega}}_{p}^{av}$,
while the main contribution is due to the final evolution when the pairs
oscillate almost with the same frequency close to ${\tilde{\omega}}_{p}^{av}$.

The upper limit to the optical depth to pair annihilation into photons is
obtained, showing that it never exceeds unity for $E<45E_{c}$.

\end{document}